\documentclass[aps,pre,onecolumn,superscriptaddress,showpacs]{revtex4-1}
\usepackage{graphicx}
\usepackage{dcolumn}
\usepackage{bm}

\usepackage{natbib}
\usepackage[
text={7in,9.5in},centering,
]{geometry}

\usepackage{float}
\usepackage{epsfig}
\usepackage{amsmath}
\usepackage{amssymb}
\usepackage{textcomp}
\usepackage{layouts}

\begin{document}

\title{Local charge conservation law as a source of gauge condition in quantum electrodynamics}

\author{Natalia Gorobey}
\affiliation{Peter the Great Saint Petersburg Polytechnic University, Polytekhnicheskaya
29, 195251, St. Petersburg, Russia}

\author{Alexander Lukyanenko}\email{alex.lukyan@mail.ru}
\affiliation{Peter the Great Saint Petersburg Polytechnic University, Polytekhnicheskaya
29, 195251, St. Petersburg, Russia}

\author{A. V. Goltsev}
\affiliation{Ioffe Physical- Technical Institute, Polytekhnicheskaya
26, 195251, St. Petersburg, Russia}

\begin{abstract}
 A formulation of quantum electrodynamics is proposed, in which the local law of conservation of electric charge serves as the source of the gauge condition. The equations of motion of the gauge variable and the density of the charge distribution in space following from this law are introduced into the quantum theory as additional conditions. Along with fixing the gauge, the interaction of charges in the modified quantum theory is described by the dynamics of the charge distribution density. The asymptotic states of free particles at spatial infinity are replaced by the initial and final states of the electromagnetic system in the form of charged wave packets.
\end{abstract}


\maketitle

\section{introduction}
When solving problems in singular dynamical theories with a continuous group of internal symmetry \cite{1}, there is a need for additional gauge conditions. In the classical theory, the choice of the gauge condition is determined by considerations of convenience and simplicity of the solution in a particular problem \cite{2}. In quantum theory, we are already talking about eliminating the divergence in the Feynman functional integral that arises during integration over the orbit of the gauge group \cite{3}, and the gauge conditions should effectively solve this problem. In this case, the calculation result should not depend on the choice of gauge conditions, which is ensured by introducing the Faddeev-Popov determinant in the measure of the Feynman integral \cite{3}. This construction of the covariant quantum theory is formalized in the framework of the BRST symmetry \cite{4,5}, and is also formulated in the Batalin-Fradkin-Vilkovyssky theorem \cite{6,7}. However, in the Yang-Mills theory, whose symmetry group is non-Abelian, anomalies arise \cite{8, 9} if the gauge conditions do not provide an unambiguous choice of the orbital element. One of the ways to overcome these difficulties may be the choice of "natural" gauge conditions, when they are equations of motion for some dynamical variables of the theory. As such, in electrodynamics, the equation of motion was proposed \cite{10}, which arises during the infinitesimal transformation of the scalar potential
\begin{equation}
 \delta{A}_0=\dot{\epsilon}.
 \label{1}
\end{equation}
In this case, the resulting equation is added to the original action as an additional condition. The result of this modification is an effective separation in the canonical form of the action of the dynamics of gauge invariant degrees of freedom and pure gauge transformations. However, an additional condition arising from variation Eq.(\ref{1}) formally violates the relativistic invariance of the theory.

In this paper, an alternative modification is proposed that arises during the gradient transformation of all components of the 4-vector of the potential of the electromagnetic field
\begin{equation}
 \delta{A}_\mu=\partial_\mu{\epsilon}.
\end{equation}
The equation of motion that follows from this variation is well known - this is the continuity equation for an electric charge
\begin{equation}
 \partial_\mu{J}^\mu=0.
 \label{3}
\end{equation}
In the simplest case considered here, when charged matter is formed by one complex scalar field $\phi$,
\begin{equation}
 J_\mu=ie[\overline{\phi}(\partial_\mu-ieA_\mu)\phi-(\partial_\mu+ieA_\mu)\overline{\phi}\phi].
\end{equation}
The continuity equation Eq. (\ref{3}) is obviously the equation of motion for the phase of the scalar field, for which we will further use the exponential form
\begin{equation}
 \phi=\rho{e^{i\alpha}}.
 \label{5}
\end{equation}
In this parametrization, we write down the Lagrangian of the original theory
\begin{equation}
L=\frac{1}{2}\int{d^3x}\left\{(\dot{A}_i-\partial_i{A_0})^2-B_i^2+\dot{\rho}^2+\rho^2[(\dot{\alpha}-eA_0)^2-(\partial_i\alpha-eA_i)^2]+V(\rho)\right\},
 \label{6}
\end{equation}
where
\begin{equation}
 B_i=[{\nabla}\times{A}]_i
 \label{7}
\end{equation}
is the magnetic field induction, as well as its modification:
\begin{equation}
 \tilde{L}=L-\int{d^3x}[e\rho^2(\dot{\alpha}-eA_0)\dot{\epsilon}-e\rho^2(\partial_i\alpha-eA_i)\partial_i\epsilon].
 \label{8}
\end{equation}
Obviously, explicit relativistic invariance is preserved in the modified theory.

Note, however, that the continuity equation Eq. (\ref{3}) is also gradient invariant. In this case, the phase is a pure gauge, in the sense that it changes additively during the gradient transformation. Thus, this modification also does not fix the gauge, but makes this procedure natural in quantum theory. This is ensured by removing the functional integration over the phase by linearizing the Hamiltonian function with respect to the corresponding canonical momentum. An additional consequence of linearization is the explicit introduction of the spatial charge distribution into the dynamics.

In the next section, the kernel of the modified evolution operator is obtained. In the second section, boundary states are defined in the form of charged wave packets. In the third section, the phase of the scalar field is excluded from the representation of the modified propagator.

\section{Quantization of the modified theory}
Let's move on to the canonical form of the modified action. For comparison, we present the Hamilton function of the original theory:
\begin{equation}
H=\int{d^3x}\left\{\frac{1}{2}(E_i^2+B_i^2)+\frac{1}{2}[\pi_\rho^2+\frac{\pi_\alpha^2}{\rho^2}+\rho^2(\partial_i\alpha-eA_i)^2+V(\rho)]+A_0G\right\},
 \label{9}
\end{equation}
where
\begin{equation}
G\equiv{e\pi_\alpha-\partial_iE_i}
\label{10}
\end{equation}
is the Gaussian constraint. Here $E_i$ we denote the canonical momentum conjugated to the vector potential of the electromagnetic field $A_i$, and $(\pi_\alpha,\pi_\rho)$ - canonical momenta of the corresponding components of the scalar field. The scalar potential $A_0$ of the electric field plays the role of a Lagrange multiplier. For the modified theory, we write the canonical momenta explicitly. First, those that do not change with modification:
\begin{equation}
 E_i=\dot{A}_i-\partial_iA_0, E_0=0, \pi_\rho=\dot{\rho},
 \label{11}
\end{equation}
where $E_0$ is the canonical momentum conjugate to the scalar potential $A_0$. The canonical momentum conjugate to the phase of the scalar field will change:
\begin{equation}
\pi_\alpha=\rho^2(\dot{\alpha}-eA_0-e\dot{\epsilon})\equiv{J}_0,
\end{equation}
and an additional momentum appears, conjugate to the infinitesimal shift $\epsilon$,
\begin{equation}
 P_\epsilon=-e\rho^2(\dot{\alpha}-eA_0).
\end{equation}
Taking into account these definitions, the Hamilton function of the modified theory will be written as:
\begin{equation}
\tilde{H}=\int{d^3x}\left\{\frac{1}{2}(E_i^2+B_i^2)+\frac{1}{2}[\pi_\rho^2+\frac{1}{\rho^2}(-2\pi_\alpha\frac{P_\epsilon}{e}-\frac{P_\epsilon^2}{e^2})+\rho^2(\partial_i\alpha-eA_i)^2+V(\rho)]+e\rho^2(\partial_i\alpha-eA_i)\partial_i\epsilon+A_0G\right\}.
 \label{14}
\end{equation}
The kernel of the unitary evolution operator corresponding to this Hamiltonian function is determined by the functional integral on the phase space of the modified theory \cite{3}. In this case, in the canonical pair $(\epsilon,P_\epsilon)$, we will consider as a momentum $\epsilon$. The functional integral over momentum variables $\pi_\alpha$ and $\epsilon$ leads to the appearance of two $\delta$-functions:
\begin{equation}
\delta(1)\equiv\delta(\dot{\alpha}-eA_0+\frac{P_\epsilon}{e\rho^2}),
\label{15}
\end{equation}
\begin{equation}
\delta(2)\equiv\delta(-\dot{P}_\epsilon+\partial_i(e\rho^2(\partial_i\alpha-eA_i))),
\label{16}
\end{equation}
which together form a pair of canonical equations of motion, equivalent to the local law of conservation of electric charge Eq. (\ref{3}). Note that a new variable has been added to the configuration space of the modified theory - the electric charge density $P_\epsilon$. The final representation of the kernel of the modified evolution operator is:
\begin{equation}
 \tilde{K}=\int \prod{dd^3A \rho d\rho} d\alpha dP_\epsilon \delta(1) \delta(2) \exp[\frac{i}{\hbar}\int_0^Tdt L(A_i,\dot{A}_i, \alpha, \dot{\alpha}, \rho, \dot{\rho})].
 \label{17}
\end{equation}
For given initial values of the phase $\alpha_0$ and density of the electric charge $P_{\epsilon{0}}$, the $\delta$-functions make it possible to remove the functional integration over these variables. Fixing the phase of the scalar field, using the corresponding equation of motion, was the purpose of the modification. However, now it is necessary to additionally associate a new dynamic variable $P_\epsilon$ with the observations. Its physical significance is revealed in the definition of the initial and final states associated with the kernel by the unitary evolution operator Eq. (\ref{17}).

\section{Charged wave packets}
When describing processes in quantum electrodynamics by the modified evolution operator Eq. (\ref{17}), due to the appearance of a new dynamic variable $P_\epsilon$, we are forced to abandon the use of asymptotic states of free motion of particles at spatial infinity. The movement of the aggregate charge in the process of interaction is now displayed explicitly in the form of its spatial density. As in a real experiment, this movement takes place in a finite region of space-time. The same description is required for pre-collision charge states and for reaction products in detecting devices. These states can be specified only with known instrumental errors in the form of wave packets. The parameters of these packages are limited primarily by the uncertainty principle. In relativistic quantum mechanics, before the second quantization of the wave function, the space charge distribution density is directly related to the distribution of the probability of particle localization in space to the extent that the wave field does not contain negative frequencies (for anti-particles, respectively, positive) \cite{11}. We will keep here this probabilistic interpretation for the dynamic variable $P_\epsilon$, supplementing it with a description of the state of the radial component of the second-quantized scalar field. In accordance with the structure of the propagator Eq. (\ref{17}), we will seek an asymptotic solution to the modified Schr\"{o}dinger equation for $t \rightarrow 0$
\begin{equation}
 i\hbar\frac{\partial\Psi}{\partial{t}}=\hat{\tilde{H}}\Psi
 \label{18}
\end{equation}
with certain values of the initial phase $\alpha_0$ and initial charge distribution density $P_{\epsilon{0}}$ in the following form:
\begin{equation}
\Psi[t,A_i,\alpha,\rho]=\delta(\alpha-\alpha(t))\delta(P_\epsilon-P_{\epsilon}(t))exp[-\frac{i}{e}\int{d^3x}\alpha P_{\epsilon{0}}-\frac{i}{\hbar}Wt]\psi[\alpha_o,P_{\epsilon{0}},A_i,\rho],
\label{19}
\end{equation}
where $\psi$ is the solution of the stationary Schr\"{o}dinger equation
\begin{equation}
\int{d^3x}\left\{\frac{1}{2}(\hat{E}_i^2+B_i^2)+\frac{1}{2}[\hat{\pi}_\rho^2+\frac{P_{\epsilon{0}}^2}{e^2\rho^2}+\rho^2(\partial_i\alpha_0-eA_i)^2+V(\rho)]+A_0\hat{G}\right\}\psi=W\psi
 \label{20}
\end{equation}
with the additional condition
\begin{equation}
\langle\psi|\hat{G}|\psi\rangle=0
\end{equation}
determining the scalar potential $A_0$. The energy operator in Eq. (\ref{20}) is actually the Hamilton operator of the original theory, in which the phase of the scalar field and the canonically conjugate momentum, the space charge density, are fixed. In this asymptotic formula, the time dependence of the phase $\alpha(t)$ and the charge distribution density $P_\epsilon(t)$ under the $\delta$-function sign is allowed, and, $\alpha(0)=\alpha_0. P_\epsilon(0)=P_{\epsilon{0}}$ Wave function Eq. (\ref{19}) is not an exact solution to the Schr\"{o}dinger equation Eq. (\ref{18}). However, it is easy to check that it is fulfilled "on average"
\begin{equation}
 \langle\psi|i\hbar\frac{\partial\Psi}{\partial{t}}-\hat{\tilde{H}}\Psi\rangle=0
\end{equation}
under additional conditions that determine the initial values of the speeds:
\begin{equation}
\dot{\alpha}_0=eA_0-\frac{P_{\epsilon{0}}}{e}\langle\psi|\frac{1}{\rho^2}|\psi\rangle,
 \label{23}
\end{equation}
\begin{equation}
 \dot{P}_{\epsilon{0}}=\langle\psi|\partial_i(e\rho^2(\partial_i\alpha_0-eA_i))|\psi\rangle.
\end{equation}
Thus, we have obtained the definition of the initial state of the system for the modified evolution operator Eq. (\ref{17}) with given definite values of the phase $\alpha_0$ and density of the charge distribution $P_{\epsilon{0}}$, if we put $t=0$ everywhere in Eq. (\ref{19}). The final state is defined similarly. For many-particle states, the space charge density has a carrier formed by a corresponding set of disjoint regions. As a consequence of the requirement that the exponential factor in Eq. (\ref{19}) be unambiguous, the total charge in each region is an integer multiple $e$. Identifying an individual charged spot in the final state with a detected charge, we must supplement this interpretation with the assumption of an ideal detector complex: each detector fires when it absorbs a whole charge, otherwise it does not fire at all. The dimensions of the detectors filling the entire space in the vicinity of the interaction region are assumed to be small compared to the dimensions of the initial wave packets. Thus, the reaction products are recorded as point particles. Excitations of an electromagnetic field - photons in the initial and final states are described in the usual way by its transverse components.

\section{Regularization of modified propagator}
Propagator Eq. (\ref{17}) remains singular, since the gradient invariance with the appearance of factors Eqs. (\ref{15}) and (\ref{16}) is not violated. However, functional integration over the phase can now be removed. With functional integration over the electromagnetic field variables, the obtained solution for the phase "runs" along all trajectories in the orbit of the gauge group between two boundary points. We make the functional integral finite if we fix this solution on an arbitrary trajectory. In this case, the multiplier Eq. (\ref{15}) should be used to remove the functional integration over the scalar potential $A_0$. Now, using the gradient transformation of the vector potential $A_i$, one can finally make the phase equal to zero everywhere, including at the boundary points corresponding to the boundary states Eq. (\ref{20}). After that, the regularized propagator takes the form:
\begin{equation}
 \tilde{K}_r=\int\prod{dd^3A\rho{d\rho}}{dP_\epsilon}\delta_r(2)exp[\frac{i}{\hbar}\int_0^TdtL_r(A_i,\dot{A}_i,P_\epsilon,\rho,\dot{\rho})],
 \label{25}
\end{equation}
where
\begin{equation}
L_r(A_i,\dot{A}_i,P_\epsilon,\rho,\dot{\rho})=\frac{1}{2}\int{d^3x}\left\{[\dot{A}_i-\partial_i(\frac{P^2_\epsilon}{e^2 \rho^2})]^2-B_i^2+\dot{\rho}^2+\frac{P^2_\epsilon}{e^2}-e^2\rho^2 A_i^2+V(\rho)\right\},
\end{equation}
and
\begin{equation}
\delta_r(2)\equiv\delta(\dot{P}_\epsilon+e\partial_i(\rho^2 A_i)).
\end{equation}
In this case, in the initial Eq. (\ref{20}) and final states, one should also set $\alpha_0= 0$.

\section{Conclusions}
Using the local law of conservation of electric charge to introduce a natural gauge condition leads to a significant change in the description of fundamental processes in quantum electrodynamics. The movement of charges in these processes, as well as their initial and final states, should be set by the local density of the charge distribution in space. The initial and final distributions of this density in the form of wave packets formed by a quantized complex scalar field specify the position and motion of charged particles with real instrumental uncertainty. Such a description can be considered as a variant of the semiclassical approximation in the quantum theory of a scalar field. However, evaluating the modification proposed here in comparison with its nonrelativistic version considered in \cite{10}, a different conclusion can be drawn. There, the modification of the original theory results in an exact separation of the dynamics of gauge invariant degrees of freedom and gauge transformations. The resulting additional dynamic variable $P_\epsilon$ is interpreted as a static charge distribution in the laboratory, which does not fundamentally change the quantum dynamics of the system. In the new, relativistically invariant version of the modified theory, the additional dynamic variable becomes the main physical parameter of the charge state of the system, and we accept such detail as acceptable at all stages of interaction. This complicates the description of fundamental processes. On the other hand, the rejection of the concept of a point charge, as well as the effective localization of processes in finite regions of space-time, eliminate the obvious sources of infinities that were present in the original theory. The proposed formalism can be generalized to non-Abelian gauge theories.

\section{Acknowledgement}

The authors thank V. A. Franke for useful discussions.

\end{document}